# Terahertz Emission From an Exchange-Coupled Synthetic Antiferromagnet

Qi Zhang,[1,2] Yumeng Yang,[2,3] Ziyan Luo,[2] Yanjun Xu,[2] Rongxiang Nie,[4] Xinhai Zhang,[1,†] and Yihong Wu[2,*]

[1] *Department of Electrical and Electronic Engineering, Southern University of Science and Technology, Xueyuan Rd 1088, Shenzhen 518055, China*

[2] *Department of Electrical and Computer Engineering, National University of Singapore, 4 Engineering Drive 3, Singapore 117583, Singapore*

[3] *School of Information Science and Technology, ShanghaiTech University, 393 Middle Huaxia Road, Shanghai 201210, China*

[4] *NUS (Suzhou) Research Institute, National University of Singapore, No. 377 Linquan Street, Suzhou 215123, China*

We report on terahertz emission (THz) from FeMnPt/Ru/FeMnPt and Pt/CoFeB/Ru/CoFeB/Pt synthetic antiferromagnetic (SAF) structures upon irradiation by a femtosecond laser; the former is via anomalous Hall effect, whereas the latter is through inverse spin Hall effect. The antiparallel alignment of the two ferromagnetic layers leads to a THz emission peak amplitude which is almost doubled as compared to the respective single layer or bilayer emitter with the same equivalent thickness. In addition, we demonstrate by both simulation and experiment that the THz emission provides a powerful tool to probe the magnetization reversal process of individual ferromagnetic layers in the SAF structure as the THz signal is proportional to the vector difference ($\mathbf{M}_1 - \mathbf{M}_2$) of the magnetization of the two ferromagnetic layers.

Corresponding authors: Email: *elewuyh@nus.edu.sg; †zhangxh@sustc.edu.cn

## I. INTRODUCTION

Recently spin-to-charge conversion in femtosecond (fs) laser excited magnetic heterostructures has attracted attention as a promising mechanism for producing terahertz (THz) emission with high efficiency, wide bandwidth and magnetically controllable polarization state [1]. The key to the THz emission is the generation of spin-polarized superdiffusive charge current from a ferromagnetic layer by fs laser excitation and subsequent conversion of the spin current to a transverse charge current, and thereby generating the THz emission. The two widely studied spin-to-charge conversion mechanisms are inverse spin Hall effect (ISHE) [1-12] and inverse Rashba-Edelstein effect (IREE) [13-15]. The former involves a ferromagnet (FM) / non-magnet (NM) heterostructure wherein when a fs laser is irradiated on the FM/NM heterostructure, non-equilibrium electrons are excited to the states above the Fermi level, generating a spin polarized superdiffusive current flowing from FM to the NM layer [16,17]. The superdiffusive transient current is subsequently converted to a transverse charge current in the NM layer with large spin-orbit coupling (SOC) and thus gives rise to the THz emission [1,2]. On the other hand, the THz emitter based on IREE typically consists of an FM with an adjacent Rashba interface, e.g., FM/Ag/Bi [18]; in this case, the superdiffusive spin polarized current launched by the FM layer is converted to transverse charge current at the Ag/Bi interface via the IREE [13-15], which in the same way as ISHE, generates the THz emission. Recently, we have demonstrated an alternative way to generate terahertz emission from a single layer FM which involves the generation of backflow nonthermal charge current from the ferromagnet/dielectric interface by fs laser excitation and subsequent conversion of the charge current to a transverse transient charge current via the anomalous Hall effect (AHE), thereby generating the THz radiation [19]. The THz emission can be either enhanced or suppressed, or even the polarity can be reversed, by introducing a magnetization gradient in the thickness direction of the ferromagnet. Unlike spintronic THz emitters based on ISHE or IREE, the AHE-based emitter does not require additional non-magnetic layer or Rashba interface.

In this work, we investigate the THz emission from exchange-coupled synthetic antiferromagnet (SAF). Two types of SAF structures have been studied, namely, FeMnPt/Ru/FeMnPt and Pt/CoFeB/Ru/CoFeB/Pt. The former is based on the AHE and the latter on the ISHE to generate THz emission. The atomic compositions of FeMnPt and CoFeB are $(Fe_{0.8}Mn_{0.2})_{0.67}Pt_{0.33}$ and $Co_{0.2}Fe_{0.6}B_{0.2}$, respectively, unless otherwise specified. The motivations are twofold. First, an enhancement of THz emission is anticipated in both cases when the magnetizations of the two FM layers are aligned in antiparallel; this is because in both structures the longitudinal superdiffusive currents would flow in opposite directions in the two FM layers. Second, since the polarization of THz emission is determined by the direction of both the superdiffusive current and the magnetization, the THz emission may provide a powerful tool to probe the magnetization reversal process in SAF - an important constituent for all types of spintronic devices. Experimentally, indeed we found that there is a significant enhancement of THz emission in the SAF structures as compared to single layer FeMnPt or CoFeB/Pt bilayer emitter with the same equivalent thickness (nearly doubled). Through both simulation and experiments, we show that the polarization of THz wave provides additional insight into the magnetization reversal process of SAF which cannot be obtained from standard magnetometry measurements. This is because the superdiffusive currents function effectively as two independent probes to the two FM layers instead of a single probe by the pump laser. The rest of the paper is organized as follows. Section II deals with macro-spin modelling of THz emission from the SAF structures, in which we will show how the THz polarization is dependent on the magnetization directions of the individual FM layers. The experimental details are given in Section III. In Section IV, we present the experimental results on both types of SAF structures, including their magnetic properties and characteristics of THz emission, based on which conclusions will be drawn for this work.

## II. MACRO-SPIN MODELLING OF TERAHERTZ EMISSION FROM SAF STRUCTRUES

Before we discuss the experimental details we first simulate the THz emission from the SAF using a macro-spin model and compare the results with the hysteresis curve obtained from magnetometry measurement. Figures 1(a) and 1(b) show the structure of the two types of SAFs, respectively. Also shown are the coordinate systems (Fig. 1(c)) which will be used for the discussion throughout this paper. Here, the films are in the *xy* plane and both the excitation laser and THz wave propagate in the *z*-direction. Although both samples have a SAF structure, the key difference is that the first structure (hereafter we refer it to as FeMnPt SAF) does not have any heavy metal layers at the two sides, whereas the second structure consists of an SAF sandwiched by two Pt layers (hereafter we refer it to as CoFeB SAF). Upon fs laser irradiation, spin polarized superdiffusive current is generated in both the FeMnPt and CoFeB layers. In the case of FeMnPt SAF, the superdiffusive current is reflected back from both the left (with quartz substrate) and right (with MgO capping layer) interfaces, which is subsequently converted to transverse charge current via AHE to generate the THz emission. Since the backflow currents from the two interfaces have opposite directions, the net transverse charge current density can be written as $\mathbf{j}_c \propto (\mathbf{M}_1 - \mathbf{M}_2) \times \hat{\mathbf{z}}$, where $\hat{\mathbf{z}}$ is a unit vector along *z*-axis. On the other hand, for the CoFeB SAF, the superdiffusive currents generated in the two FM layers enter their respective adjacent Pt layers where they are converted to transverse charge current via ISHE, which eventually emits as a THz wave. Similar to the case of FeMnPt SAF, the net transverse charge current density can be written as $\mathbf{j}_c \propto -(\mathbf{M}_1 - \mathbf{M}_2) \times \hat{\mathbf{z}}$. The difference, if any, is just in the sign which depends on the relative magnitudes of the superdiffusive currents in the two layers as well as the transparency of the interfaces which are not always the same. These simple facts show that THz generation of the two SAF structures can be modeled using the macro-spin model on an equal footing, despite that they are based on different mechanisms. The amplitude of the THz wave detected by the detector depends on the relative orientation of the magnetization of the FM layers, the polarization axis of the analyzer, and the crystal direction of the ZnTe detector. Without losing generality, we assume that the polarization axis of the wire grid analyzer and the detection axis of the ZnTe detector are at angles of $\varphi_a$ and $\varphi_d$ [20], respectively, away from the y-axis (Fig. 1(c)). Similarly, the directions of $\mathbf{M}_1$ and $\mathbf{M}_2$ are defined by $\varphi_1$ and $\varphi_2$, respectively (assuming both are in the plane). We use the convention that $\varphi_1$ and $\varphi_2$ are positive when they rotate away from positive y- to negative x-axis (or counterclockwise rotation) and negative when they rotate in the opposite direction. By using this sign convention and the coordinate system in Fig. 1(c), the detected electric field of the THz emission may be written as

$$\mathbf{E}_{THz} \propto \pm[(M_1 \cos\varphi_1 - M_2 \cos\varphi_2)\sin\varphi_a \sin\varphi_d, (M_2 \sin\varphi_2 - M_1 \sin\varphi_1)\cos\varphi_a \cos\varphi_d, 0].$$

(1)

Here we assume that the spin polarized superdiffusive current is proportional to the magnetization. Although there is no rigorous basis for this assumption, it won't affect the discussion because we are only interested in its direction instead of absolute value. The sign before the square brackets depends on the direction of spin-polarized current that contribute THz emission in two SAFs, i.e., positive for the FeMnPt SAF and negative for the CoFeB SAF. If we let $\varphi_a = \varphi_d = \pi/2$, then $\mathbf{E}_{THz}$ would only have the x component which is given by $E_x \propto \pm(M_1 \cos\varphi_1 - M_2 \cos\varphi_2) = \pm(M_{1y} - M_{2y}) = \pm\Delta M_y$. Similarly, when $\varphi_a = \varphi_d = 0$, there is only a y component given by $E_y \propto \pm(M_2 \sin\varphi_2 - M_1 \sin\varphi_1) = \pm(M_{2x} - M_{1x}) = \mp\Delta M_x$. It is apparent that the THz emission will be maximum when the two layers are antiparallel to each other. The amplitude of the THz wave will be simply the sum of the contributions from the two layers, which means that theoretically it could be doubled as compared to the single layer emitter of same equivalent thickness if the effect of Ru is small. The latter is a reasonable assumption because i) the superdiffusive currents from the two sides of Ru are presumably cancelled out with each other and ii) Ru has a relative small spin Hall angle compared with other heavy metals used in spintronic THz emitters [2,12,21,22]. Moreover, the superdiffusive current crosses over the Ru layer entering the other side of the FeMnPt creates a desirable effect because it increases the difference in backflow current from the FM/oxide and FM/Ru interface. The simple relationship we derived above may serve as a powerful tool to study the magnetization reversal process of SAF when it is subject to a sweeping magnetic field. By examining the sign of $E_x$ and $E_y$, one would be able to find out in which direction and how the magnetizations of the two FM layers would respond to the sweeping external field. The information obtained will be very useful in gaining an insight of exchange coupling in the SAF, which is important for device applications. In contrast, the standard magnetometry measurement can only provide information about the overall projection of magnetization of the two layers, i.e., $M_y = M_1 \cos\varphi_1 + M_2 \cos\varphi_2$; it is difficult to probe the magnetization directions of the individual layers which are crucial for understanding the magnetization reversal process.

To have a more quantitative understanding of the difference between the information obtained by the THz and magnetometry measurements, we firstly conduct a macro-spin modeling of the SAF structure. For a SAF consisting of two magnetic layers with uniaxial magnetic anisotropy, in the coordinate system of Fig. 1(c), the energy per unit area as a function of an in-plane magnetic field ($H$) is given by

$$E = K_{u1} d_1 \sin^2\varphi_1 + K_{u2} d_2 \sin^2\varphi_2 - H\left[d_1 M_1 \cos(\varphi_H - \varphi_1) + d_2 M_2 \cos(\varphi_H - \varphi_2)\right] + J_1 \cos(\varphi_1 - \varphi_2) + J_2 \cos^2(\varphi_1 - \varphi_2)$$

(2)

where $K_{u1}$ and $K_{u2}$ are the uniaxial anisotropy constant, $\varphi_H$ is the angle between $H$ and the SAF easy axis (i.e., $y$-axis), $M_1(M_2)$ and $d_1(d_2)$ are the saturation magnetization and thickness of the two FM layers, respectively, $J_1$ and $J_2$ are the bilinear and biquadratic exchange coupling constant [23].

A positive $J_1$ favors antiparallel coupling [24], leading to a SAF structure. The biquadratic coupling constant, $J_2$, is introduced to account for the effect of spacer pinhole and interface roughness [25]. In order to be consistent with the measurement configuration, hereafter we set $\varphi_H = 0$, i.e., we only sweep the field along *y*-axis. The simulated results of $\varphi_1$ and $\varphi_2$ as a function of *H* for both forward (solid-line) and backward (dashed-line) sweeping of the structures are given in Appendix A. As discussed in detail in Appendix A, depending on the sense of rotation of the layer which starts to rotate first, there are four possible $\varphi_1$ and $\varphi_2$ combinations during a complete loop of field sweeping, which will lead to different types of field-dependence of THz emission, as discussed below.

Based on the $\varphi_1$ and $\varphi_2$ values given in Fig. 7 in Appendix A, we can calculate $M_y$, $\Delta M_x$, and $\Delta M_y$ as a function of external field *H*. As we discussed above, the $M_y$-*H* curve corresponds to the *M-H* loop measured by VSM, and $\mp\Delta M_x$-*H* and $\pm\Delta M_y$-*H* are corresponding to $E_y$ and $E_x$ of the THz wave as a function of *H*. The calculation results for the FeMnPt SAF are shown in Figs. 2(a)-(d), Figs. 2(e)-(h), and Figs. 2(i)-(l) for $M_y$, $\Delta M_y$, and $-\Delta M_x$, respectively. Each set of these curves corresponds to the $\varphi_1$ and $\varphi_2$ results shown in Figs. 7(a)-(d) in Appendix A. As can be seen from the figures, both the $M_y$-*H* and $\Delta M_y$-*H* curves cannot tell the difference among the four different types of magnetization reversal process. However, the $-\Delta M_x$-*H* loops are distinct in the four different cases. As we will show later experimentally that this difference serves as a powerful tool to probe the magnetization reversal process in SAF structures. Similar results are obtained for the CoFeB SAF as shown in Figs. 3[(a)-(d): $M_y$; (e)-(h): $-\Delta M_y$; (i)-(l): $\Delta M_x$]. As with the case of FeMnPt SAF, there is no difference among the four different cases for $M_y$ and $\Delta M_y$, but there is a clear difference in $-\Delta M_x$. Since the absolute polarity of the THz emission can be obtained from control samples with a single layer FM, the four types of magnetization reversal process can be unambiguously distinguished by the polarity of $E_y$ of the THz wave.

## III. EXPERIMENTAL DETAILS

The samples investigated in this study are listed in Table I. All the samples are deposited on 500-μm-thick fused quartz substrates using DC magnetron sputtering with a base and working pressure of $2\times10^{-8}$ Torr and $3\times10^{-3}$ Torr, respectively. The FeMnPt films are prepared by co-sputtering of $Fe_{0.8}Mn_{0.2}$ and Pt targets. The chemical composition of all the samples is determined by X-ray photoelectron spectroscopy (XPS) [26]. With the $Fe_{80}Mn_{20}$ and Pt cathode power at 50 W and 15 W, respectively, we are able to produce a $(Fe_{0.8}Mn_{0.2})_{0.67}Pt_{0.33}$ film with an anomalous Hall angle of $\theta_{AH} = 0.0269$, which is comparable to the spin Hall angle of Pt [26]. For simplicity, hereafter, we refer $(Fe_{0.8}Mn_{0.2})_{0.67}Pt_{0.33}$ to as FeMnPt unless otherwise specified. A 4 nm MgO capping layer is deposited on top of the FeMnPt to prevent it from oxidation [27]. As discussed above, it also functions as a reflective layer for superdiffusive electrons. The CoFeB is deposited from a single target of $Co_{20}Fe_{60}B_{20}$. A standard THz emission spectroscopy system is used for generation and detection of THz wave from different samples. The system is driven by a femtosecond Ti:sapphire laser with a pulse width of 35 fs, repetition rate of 1 kHz, wavelength of 800 nm and excitation laser power of 120 mW (equivalent to a fluence of 330 μJ/cm$^2$). During field sweeping measurement, an in-plane magnetic field is applied in the easy axis direction using an electromagnet. In our particular measurement configuration, the easy axis is aligned along *y*-axis, and *z*-axis is the laser and THz wave propagation direction. The THz emission is calibrated [28] and detected using electric-optical (EO) sampling with a 1000 μm thick ZnTe with <110> crystalline orientation, where the magnitude in all figures indicate the EO signal acquired by the balanced detector. All the measurements are performed in a dry nitrogen gas environment. The magnetometry measurements are carried out using the Quantum Design's VersaLab system.

TABLE I. Samples used in this study. The number inside the brackets denotes the thickness in nanometer. FeMnPt and CoFeB refer to $(Fe_{0.8}Mn_{0.2})_{0.67}Pt_{0.33}$ and $Co_{20}Fe_{60}B_{20}$, respectively.

| Sample Name | Sample Structure |
|---|---|
| FeMnPt SAF | Quartz/FeMnPt(4)/Ru(0.8)/FeMnPt(4)/MgO(4) |
| CoFeB SAF | Quartz/Pt(2)/CoFeB(4)/Ru(0.8)/CoFeB(3)/Pt(2)/MgO(4) |
| FeMnPt single-layer | Quartz/FeMnPt(5)/MgO(4) |
| CoFeB/Pt bilayer | Quartz/CoFeB(3)/Pt(3)/MgO(4) |

## IV. RESULTS AND DISCUSSION

### A. Magnetic properties of FeMnPt and CoFeB SAFs

We firstly study the magnetic properties of the FeMnPt and CoFeB SAFs using the VSM with magnetic field applied parallel to the easy axis direction. Figure 4(a) shows the *M-H* curves of five FeMnPt SAFs with the structure of quartz/FeMnPt($d_1$)/Ru(0.8)/FeMnPt($d_2$)/MgO(4). The curves are shifted vertically for clarity. For simplicity, we use ($d_1$, $d_2$) to indicate the thickness combination (in nm) of the two FeMnPt layers in the SAF. Except for the (3, 2) sample, all other SAFs exhibit the antiparallel state of $\mathbf{M}_1$ and $\mathbf{M}_2$ at small field. In what follows we take the (9, 3) sample as an example to explain briefly how the magnetizations respond to a sweeping external field. In order to be consistent with the simulation results discussed above, we begin with the backward sweeping process first. At the initial stage, both $\mathbf{M}_1$ and $\mathbf{M}_2$ are aligned along the +*y* direction. The net magnetization in the filed direction is $M_1 + M_2$. When the field strength is gradually reduced from 200 Oe, $\mathbf{M}_2$ is reversed at around 95 Oe due to the negative exchange coupling field. This leads to the recovery of the first AF state with a net magnetization of $M_1 - M_2$. The AF state is maintained until the field is reduced to zero and then changes its direction. Upon further increasing the field strength in the negative direction, $\mathbf{M}_1$ starts to rotate to opposite direction at –8 Oe due to the Zeeman energy gain, and the magnetization reversal process is completed at about 24 Oe. This will induce a simultaneous flip of $\mathbf{M}_2$ back to the $+y$ direction, thereby leading to the second AF state with a net magnetization of $M_2 - M_1$. When the field strength is further increased in the negative direction to –95 Oe, both $\mathbf{M}_1$ and $\mathbf{M}_2$ will be aligned along the $-y$ direction, which gives a net magnetization of $-M_1 - M_2$. The reversals of magnetization will take place during forward sweeping from –200 Oe to 200 Oe. The reversals of magnetization of the (5, 3) and (4, 3) samples are essentially the same as those of (9, 3) except that the field at which $\mathbf{M}_2$ reverses becomes smaller and the net magnetization also changes more gradually with the external field due to the rotation of $\mathbf{M}_1$ as well. However, a clear difference is seen in the *M-H* loop of the (4, 4) sample. Unlike the other 3 samples, $\mathbf{M}_2$ stays in the $-y$ direction after it is switched at a positive field. The flipping of $\mathbf{M}_2$ to $+y$ direction doesn't happen at negative field because it does not result in a significant Zeeman energy gain by simply switching the direction of $\mathbf{M}_1$ and $\mathbf{M}_2$ when the difference between $M_1 d_1$ and $M_2 d_2$ is small. Note that although the two FM layers have the same thickness in the (4, 4) sample, their magnetization and thickness products are slightly different due to the fact that the two layers are deposited on different materials, one directly on the quartz substrate and the other one on the Ru layer. For the sake of simplicity, we assume that the actual thickness is the same as the nominal thickness, and the difference between $M_1 d_1$ and $M_2 d_2$ is mainly due to the difference in magnetization. This gives the saturation magnetization of $M_1 = 262$ emu/cm$^3$ and $M_2 = 198$ emu/cm$^3$. These values are used in the simulation. As can be seen, the *M-H* loop obtained from the VSM measurement is in good agreement with the simulation results shown in Fig. 2(a). In addition, we also measured the *M-H* loops of the (4, 4) sample at elevated temperatures from 300 K to 400 K and the results are shown in Fig. 4(b). The *M-H* loops at different temperatures are quantitatively the same, though the switching field becomes smaller and the switch itself also becomes sharper when the temperature increases. This is presumably caused by the decrease of both exchange constant and uniaxial anisotropy of the FeMnPt layers induced by temperature. Despite this, the SAF state is maintained even at a temperature up to 400 K.

The *M-H* loops for the four CoFeB SAFs with a thickness combination of (6, 6), (5, 5), (4, 3) and (4, 2.5) are shown in Fig. 4(c). The inset is the enlarged portion of the low field region. As discussed in the simulation section, the large $J_1$ and $J_2$ values result in a more gradual rotation of $\mathbf{M}_1$ and $\mathbf{M}_2$ as compared to the FeMnPt SAFs. The experimental *M-H* loops are in qualitative agreement with the simulated ones shown in Figs. 3(a)-(d). As shown in the inset, among the four samples, the (4, 3) sample gives the smallest net moment at zero field, and therefore, in both the simulation and THz measurement, we focus on this sample. As can be seen, a field up to 4000 Oe is required to saturate $\mathbf{M}_1$ and $\mathbf{M}_2$ in the +*y* direction. During backward sweeping, when the magnetic field decreases from 4000 Oe, $\mathbf{M}_1$ and $\mathbf{M}_2$ start to rotate gradually in opposite direction, leading to a gradual decrease of the magnetization from 680 emu/cm$^3$ at 4000 Oe to 80 emu/cm$^3$ at 100 Oe. At this field, $\mathbf{M}_2$ is nearly reversed, whereas $\mathbf{M}_1$ goes back to original saturation direction along +*y* axis. This corresponds to the first AF state, from which we can infer that $M_1 = 380$ emu/cm$^3$ and $M_2 = 300$ emu/cm$^3$, assuming the actual thickness is the same as the nominal thickness. Shortly after the field is reversed, we reach the second AF state in which both $\mathbf{M}_1$ and $\mathbf{M}_2$ switch its direction after which both gradually rotate and saturate along the –*y* direction at around –4000 Oe. The forward sweeping shows the similar trend, which eventually leads to the experimentally observed *M-H* loop. Figure 4(d) shows the *M-H* loops for the same sample at temperatures from 300 K to 400 K. Again, the inset shows the zoom-in view of the low field region. When the temperature increases, it becomes easier to saturate the magnetization in both field directions due to the decrease of exchange constant (4000 Oe at 300 K versus 2430 Oe at 400 K). In the meantime, the field range for the AF state also becomes narrower. As with the case of the (4, 4) FeMnPt sample, the (4, 3) CoFeB sample also maintains its SAF configuration at 400 K. This ensures that the sample can sustain the heating effect, if any, from the laser irradiation during the THz measurement.

## B. THz emission of FeMnPt and CoFeB SAFs

We now turn to the THz emission from the (4, 4) FeMnPt and (4, 3) CoFeB SAFs. Figure 5(a) shows the THz time domain waveforms of the (4, 4) FeMnPt sample at an applied field of 0 (dashed-line) and 200 Oe (solid-line), respectively. The laser fluence is 330 μJ/cm$^2$. For comparison, we also show the result for a single-layer FeMnPt emitter with a thickness of 5 nm (dotted-line) measured under the same condition. It should be pointed out the time-domain waveforms shown in Fig. 5(a) and (b) have been normalized by the laser fluence. The peak amplitude ratios are as follows: $E_{SAF(0\ Oe)} / E_{SAF(200\ Oe)} = 3.99$ and $E_{SAF(200\ Oe)} / E_{FeMnPt} = 0.55$. As we demonstrated previously [19], in the case of FeMnPt single layer emitter, the THz amplitude is determined by the net longitudinal superdiffusive current $j_l = j_{l1} - j_{l2}$, where $j_{l1}$ and $j_{l2}$ are the backflow spin polarized current from the two interfaces, one with the substrate and the other with MgO capping layer. We have also shown that the THz emission is strongly dependent on the FeMnPt thickness. The FeMnPt SAF shown in Fig. 1(a) has four interfaces: (i) quartz/FeMnPt, (ii) FeMnPt/Ru, (iii) Ru/FeMnPt, and (iv) FeMnPt/MgO. Since both the FeMnPt and Ru are metallic layers, we may assume that the backflow currents from interfaces (ii) and (iii) are much smaller than those from the interfaces with MgO and substrate. This is a reasonable assumption considering the polycrystalline nature of the sample, although further studies are required to quantify the reflection at the metallic interfaces [29]. Therefore, at 200 Oe, the SAF may be treated as a single layer emitter with an equivalent thickness of 8.8 nm. As shown in our previous work, for a single layer FeMnPt emitter, the THz emission amplitude at 8.8 nm is about 60% of that at 5 nm. This well explains the $E_{SAF(200\ Oe)} / E_{FeMnPt}$ ratio of 0.55. On the other hand, at zero field, i.e., when the two FMs are antiparallel, the contributions from the quartz/FeMnPt and FeMnPt/MgO add up instead of cancelling out with each other. Therefore, as a simple estimation, the $E_{SAF(0\ Oe)} / E_{SAF(200\ Oe)}$ ratio may be approximated as $(1+\alpha)/(1-\alpha)$, where $\alpha$ is the ratio between backflow current from the MgO and quartz interfaces. At a $E_{SAF(0\ Oe)}/E_{SAF(200}$ $_{Oe)}$ ratio of 3.99, $\alpha$ turned out to be 0.6. The difference between the top and bottom interface is not as large as what we have found previously in the case of single emitters. This is presumably caused by the insertion of a thin Ru which helps to reduce the roughness of the FeMnPt/MgO interface. This is a desirable result as it significantly enhances the THz emission at the AF state (see more in Appendix B). These results further affirm the AHE origin of the THz emission as proposed in our previous study.

Similar response to the external field was observed in the THz emission from the CoFeB except that the polarization is opposite to that of the FeMnPt SAF. Figure 5(b) shows the THz waveforms from the (4, 3) CoFeB SAF at 0 Oe and 2000 Oe, respectively. Also shown is the THz waveform from the CoFeB(3)/Pt(3) bilayer control sample with a negative applied field. Similar to the case of FeMnPt SAF, the THz emission amplitude is maximum at 0 Oe but nearly zero at saturation field. The opposite polarity at zero field can be understood as follows. When the pumping directions are the same, i.e., both are from the substrate side as shown in Fig. 1, the THz polarization of the FeMnPt SAF follows the sign of $\mathbf{M}_1$, assuming $j_{l1} > j_{l2}$. Note that whether the polarization is positive or negative is relative and it has been calibrated using the single layer sample. We have previously demonstrated that the anomalous Hall angle of FeMnPt and spin Hall angle of Pt have the same sign [26]. Therefore, under the same measurement configuration, the THz polarization of the CoFeB SAF should follow the sign of $\mathbf{M}_2$. The opposite polarity of Fig. 5(b) was obtained by first saturating the sample in +$y$ direction and then reducing the field to zero. In this case, the sample is at the first AF state, i.e., $\mathbf{M}_1$ is in +$y$ and $\mathbf{M_2}$ in –$y$ direction. Although it is not shown here, we have verified that the polarization would be opposite if we first saturate the sample using a large negative field and then reduces the field to zero. This explains why the polarity is opposite for the two samples. As can be seen in Fig. 5(b), the amplitude of the (4, 3) CoFeB SAF sample at zero field (solid-line) is about 82% of that of the CoFeB(3)/Pt(3) bilayer sample (dotted-line). But this does not mean that there is no enhancement of THz emission in the SAF sample. As demonstrated by previous studies, the THz emission of FM/HM bilayers is strongly dependent on the individual layer as well as the total thickness. The total thickness of the (4, 3) CoFeB SAF sample is 11.8 nm, almost double of that the CoFeB(3)/Pt(3) bilayer sample. If we treat the SAF sample equivalently as a bilayer sample with a total thickness of 11.8 nm, the estimated peak amplitude will be 37% of that of the bilayer sample [2,30]. The fact that the peak amplitude of the SAF sample at zero field is 82% of that of the bilayer sample implies that the THz signal of the SAF sample indeed is the sum of the contributions from the two FM layers at zero field. We have also investigated the pumping fluence dependence of the SAF samples and the results are shown in Fig. 5(c). In order to avoid damaging the sample, we only increase the laser fluence to 385 μJ/cm$^2$ for the FeMnPt SAF, and for the CoFeB SAF, we increase it to 550 μJ/cm$^2$. Both samples show a linear increase with the pumping fluence, indicating that although laser heating increases the sample temperature, it does not affect the THz emission obviously at this fluence range. The large thickness makes the SAF sample relatively more insensitive to heating effect, which is potentially an advantage of the SAF emitter compared to the single layer or bilayer emitter when it comes to practical applications.

## C. Field dependence of THz emission from FeMnPt and CoFeB SAFs

As demonstrated by the simulation results presented in Section II, the THz emission provides a convenient tool to study the magnetization reversal process of the SAF structure. In order to see how the two magnetic layers rotate during field sweeping, we fix the time at the first peak of the THz waveform and investigate how the peak amplitude varies with the external field. We have performed measurements at different delay times either before or after the peak position. The shape of the THz-H loop looks almost the same except for the amplitude, suggesting that magnetic dipole radiation has negligible effect on the THz-$H$ loops due to its small contribution [19]. Figure 6(a) shows the peak value of the THz $E_x$ component of the (4, 4) FeMnPt SAF versus the applied magnetic field at a laser fluence from 110 to 385 μJ/cm². The solid (dashed) line corresponds to forward (backward) sweeping of the magnetic field. The overall shape of the $E_x$-H curve agrees well with the simulated $\Delta M_y$-H curves shown in Figs. 2(e)-(h). When the external field is large, both $\mathbf{M}_1$ and $\mathbf{M}_2$ are aligned in the field direction, resulting in a small residual emission (negative when $H < 0$ and positive when $H > 0$). When H is within ±39 Oe, the THz amplitude is maximum due to antiparallel alignment of $\mathbf{M}_1$ and $\mathbf{M}_2$. The two AF states produce THz emission of same amplitude but opposite polarization. With increasing the laser fluence from 110 to 385 μJ/cm², although the amplitude increases almost linearly, the shape remains almost unchanged; the width of the central region doesn't show an obvious narrowing trend, which is different from the temperature dependence results, meaning that the increase of the laser fluence up to 385 μJ/cm² doesn't cause additional heating effect. This corresponds well with the linear increase of THz peak amplitude shown in Fig. 5(c). As discussed in Section II, the $E_x$ component of THz emission is unable to probe the direction of $\mathbf{M}_1$ and $\mathbf{M}_2$ during the magnetization reversal process. To this end, we have measured the field-dependence of the $E_y$ component by rotating both the analyzer and detector by 90°. Compared to the $E_x$ component, it is more challenging to the measure the $E_y$ component of the FeMnPt SAF. This is because the reversal of magnetization of the two layers occurs in a narrow magnetic field range (~24 Oe), the $E_y$ signal is much weaker than $E_x$. To mitigate the situation, we have repeated the measurements for 30 times and Fig. 6(b) shows the averaged signal. For this particular measurement, the laser fluence is 330 μJ/cm². As expected, there is a sharp peak near the switching region during both backward and forward sweeping. The square-like hysteresis loop at low field may be caused by the slight misalignment of the easy axis with the external field (~1°). The measurement starts with backward sweeping. At 200 Oe, both $\mathbf{M}_1$ and $\mathbf{M}_2$ are aligned in +y direction, and therefore, there is no $E_y$ component detected. When the field decreases to around 60 Oe, both $\mathbf{M}_1$ and $\mathbf{M}_2$ start to rotate away from the y axis, leading to a non-zero net projection of the magnetization in *x*-direction, thereby giving rise to a sharp $E_y$ peak. The sign of the $E_y$ during field sweeping is similar to the simulated result of $-\Delta M_x$ in Fig. 2(i) where a positive (negative) peak appears during the backward (forward) sweeping. Therefore, the $E_y$-$H$ curve unambiguously determines the rotation direction of the magnetization, i.e., $\mathbf{M}_1$ rotates clockwisely whereas $\mathbf{M}_2$ rotates counterclockwisely respectively during both the forward and backward sweeping for this particular sample. Now the question is: what determines the rotation direction given the fact that, from energetic point of view, there is no difference between clockwise and counterclockwise rotation? Theoretically, this is indeed the case. However, during actual measurement, there is always a possibility of misalignment between the external field direction and the easy axis direction of the sample. When the misalignment is present, the layer with a larger magnetization should deviate less from the field direction so as to minimize the Zeeman energy. This will set a preferred rotation direction for the layer with larger magnetization. Once this is set, the other layer will rotate to opposite direction to minimize the exchange energy. This explains why the particular $E_y$-$H$ curve shown in Fig. 6(b) is obtained.

We now turn to the CoFeB SAF sample. Figure 6(c) shows the peak value of the THz $E_x$ component of the (4, 3) CoFeB SAF versus the applied magnetic field at a laser fluence from 110 to 550 µJ/cm². The solid (dashed) line corresponds to forward (backward) sweeping of the magnetic field. Again THz emission does not change much with the pumping fluence except for the peak amplitude. The overall shape of the $E_x$-$H$ curve agrees well with the simulated $-\Delta M_y$-$H$ curves shown in Figs. 3(e)-(h). The gradual change of the THz emission with the external field is in good agreement with the VSM results, although the saturation field is smaller in the THz measurement. This is presumably caused by the laser heating induced reduction of the exchange constant as also observed in the FeMnPt sample. Now we explain briefly the $-\Delta M_y$-$H$ curve by focusing on the measurement done at a fluence of 330 µJ/cm². We can see that the THz amplitude gradually increases in negative direction when the field is reduced from 2000 Oe. This is because $\mathbf{M}_1$ and $\mathbf{M}_2$ start to rotate away from $+y$ axis in opposite direction. Since $M_2d_2$ is smaller than $M_1d_1$, $\mathbf{M}_2$ rotates faster than $\mathbf{M}_1$ with respect to the field change. When the angle between $\mathbf{M}_2$ and $+y$ axis reaches a critical value, $\mathbf{M}_1$ starts to rotate back to $+y$. This will further increase the speed of $\mathbf{M}_2$ approaching $-y$ direction. Therefore, the $E_x$ keeps increasing in the negative direction until it reaches the negative maximum. When the field reverses its direction and increases in strength, it is energetically more favorable to align $\mathbf{M}_1$ with the external field as $M_1d_1 > M_2d_2$. This forces $\mathbf{M}_2$ to rotate back to $+y$ direction and at the same time $\mathbf{M}_1$ rotates to $-y$ direction. As a result, the $E_x$ starts to increase at around $-80$ Oe and reaches positive maximum at $-250$ Oe. From $-250$ Oe to $-2000$ Oe, the THz emission gradually decreases to zero due to the gradual alignment of the two magnetic layers with the external field. The forward sweeping curve can be understood in the same way. As with the case of FeMnPt SAF, to probe the rotation directions of $\mathbf{M}_1$ and $\mathbf{M}_2$, we have to measure the $E_y$ component. The field-sweeping measurement results for $E_y$ is shown in Fig. 6(d). Unlike the case of FeMnPt SAF, the $E_y$ always maintains the same sign during both the backward and forward sweeping. The minimum of $E_y$ at positive and negative field correspond to the maximum of $E_x$ at positive and negative field during forward and backward sweeping. Apart from a slightly larger hysteresis, the measured $E_y$-$H$ relation resembles well the simulated results in Fig. 3(k). This suggests that, instead of rotating by one cycle in the same sense of rotation, i.e., either clockwise or counterclockwise, during the backward and forward sweeping, both $\mathbf{M}_1$ and $\mathbf{M}_2$ will rotate back when the sweeping direction is reversed. This is so because in the THz measurement, even at a field of 2000 Oe (limited by the electromagnet used in the setup), we are unable to saturate the magnetization of the CoFeB layers. This explains the difference between the FeMnPt and CoFeB SAF samples. Again, the initial rotation direction of $\mathbf{M}_1$ and $\mathbf{M}_2$ when the field is reduced from 2000 Oe is set by the misalignment angle between the applied field and easy axis.

## D. Discussion

As demonstrated above both theoretically and experimentally, by measuring the $E_x$ and $E_y$ components of the THz emission, we are able to determine the rotation direction of $\mathbf{M}_1$ and $\mathbf{M}_2$ unambiguously, which is not possible through standard magnetometry measurements. Apparently, the THz signal is more informative than the net moment because (i) it is a vector which can be used to map out the magnetization direction of each layer and (ii) it is even more sensitive when $M_1 d_1 \approx M_2 d_2$. It is worth pointing out that, as long as in-plane magnetization of an individual SAF layer is concerned, the THz-based technique is superior than other commonly used magnetic characterization techniques which are able to probe the magnetization reversal process such as magneto-optic Kerr effect (MOKE) and X-ray magnetic circular dichroism (XMCD). The key difference is that in both MOKE and XMCD, the light propagation and polarization directions are the same for the two layers. In contrast, in the case of THz measurement, although the direct excitation source is the pump laser, it is not really the probe beam *per se*. The actual probe beams are the two superdiffusive currents with opposite directions. This is equivalent to a MOKE measurement whereby two laser beams with different polarization directions are irradiated on the two FM layers of a SAF structure at the same time, which is obviously very difficult to achieve. An alternative way is to measure the longitudinal and transverse separately by re-configure the MOKE optics, but this will increase complexity of the measurement system. Even this is possible, another difficult situation for MOKE is when the signals from the two FM layers have the same magnitude but opposite sign. As the net signal is zero in this case, it is impossible to determine the individual magnetization directions. Nevertheless, we do see that the combination of THz and MOKE measurements provides a powerful tool to probe magnetization dynamics of magnetic multilayers as they are largely complementary to each other. On the other hand, although the XMCD has the additional elemental selectivity, it still cannot differentiate $\mathbf{M}_1$ and $\mathbf{M}_2$ if the two layers are the same type of material, like the present case. The only disadvantage of the THz based technique at this moment is its inability to probe vertical component, which requires further investigations.

## V. CONCLUSIONS

In summary, we have demonstrated THz emission from exchange-coupled SAF structures based on both the AHE and ISHE effects. In addition to the enhancement of THz emission, we have shown both theoretically and experimentally that the THz emission is a powerful tool to probe the magnetization reversal process of SAF structures. This is possible because the THz emission from the SAF structure is proportional to the vector difference of the two magnetizations, $\mathbf{M}_1 - \mathbf{M}_2$, rather than the longitudinal component of ($\mathbf{M}_1 + \mathbf{M}_2$). The unique capability of the THz emission as a magnetization probing tool comes from the fact that the excitation laser generates two secondary probes which are the superdiffusive current with opposite directions. The separate secondary probe for each layer gives the THz emission unparalleled advantages over other conventional magnetization probing techniques such as MOKE and XMCD.

## ACKNOWLEDGEMENTS

Y.H.W. would like to acknowledge support by Ministry of Education, Singapore under its Tier 2 Grants (grant no. MOE2017-T2-2-011 and MOE2018-T2-1-076). X.H.Z. would like to acknowledge the project by Shenzhen Peacock Plan (grant no. KQTD2015071710313656) and the project of Shenzhen Science and Technology Innovation Committee (grant no. JCYJ20180504170604552).

## APPENDIX A: MACRO-SPIN SIMULATION ON SAFS

We conducted the macrospin simulation on FeMnPt and CoFeB SAFs based on Eq. (2). Figures

7(a)-(d) show the simulated results of $\varphi_1$ and $\varphi_2$ as a function of *H* for both forward (solid-line) and backward (dashed-line) sweeping of the FeMnPt SAF structure. As stated in the section II, we adopt the convention that $\varphi_1$ and $\varphi_2$ are positive for counterclockwise rotation and negative for clockwise rotation; therefore, both 180º and –180º refer to –*y* direction, and 360º and –360º are the same as 0º, which is in +*y* direction. The parameters used are $M_1$ = 262 emu/cm$^3$, $M_2$ = 198 emu/cm$^3$, $K_{u1}$ = 4 × 10$^3$ erg/cm$^3$, $K_{u2}$ = 4 × 10$^3$ erg/cm$^3$, $d_1 = d_2 = 4$ nm, $J_1 = 0.0041$ erg/cm$^2$, and $J_2 = 0$. The saturation magnetizations and thicknesses are determined experimentally. The uniaxial anisotropy constants are estimated from the relation $K_{ui} = M_i H_{ki}/2$ ($i$ = 1 or 2) where we use the coercivity of the respective layers to approximate the anisotropy field $H_k$. Once these parameters are known, $J_1$ can be calculated analytically from the field at which the first reversal occurs in the measured *M-H* loop of the SAF [31]. The final values of $K_{u1}$, $K_{u2}$, and $J_1$ are optimized by comparing the simulated *M-H* loop with the one measured by a vibrating sample magnetometer (VSM). We found that a reasonably good fitting is obtained when $J_2 = 0$. The magnetizations of the two FM layers are initially aligned antiparallel at zero field. When an external field of adequate strength is applied in +*y* direction, the two magnetizations will be aligned in the field direction (i.e., $\varphi_1 = \varphi_2 = 0$). When the field strength is gradually reduced, both $\varphi_1$ and $\varphi_2$ will start to deviate from their initial directions. Since $J_1 > 0$, **M**$_1$ and **M**$_2$ would have to rotate in opposite directions. When the external field is perfectly aligned with the easy axis direction, there is no pre-determined preference as to which one will rotate in what direction, and therefore, there are two possible senses of rotation for both the forward and backward sweeping processes, which are summarized in Table II and also illustrated in Figs. 7(i)-(l), respectively. Corresponding to the four scenarios are four sets of $\varphi_1$ and $\varphi_2$ values, which are shown in Figs. 7(a)-(d), respectively.

  We now use Fig. 7(a) as an example to explain how $\varphi_1$ and $\varphi_2$ vary with the external field starting from backward sweeping. In this case, both magnetizations are initially aligned in +*y* direction, i.e., $\varphi_1 = \varphi_2 = 0$. When the field strength is reduced, initially both $\varphi_1$ and $\varphi_2$ remain at 0°. However, when *H* decreases to 50 Oe, $\varphi_1$ ($\varphi_2$) becomes negative (positive) with its absolute value gradually increasing with decreasing the magnetic field. This means that $\mathbf{M}_1$ starts to rotate away from the *y*-axis in clockwise direction, whereas $\mathbf{M}_2$ rotates in counterclockwise direction. By further reducing the magnetic field, at around *H* = 34 Oe, $\varphi_2$ jumps abruptly to 180°, i.e., $\mathbf{M}_2$ reverses its direction. As expected, $\mathbf{M}_1$ rotates back to +*y* direction in order to minimize the energy. This is the first antiferromagnetic coupling (AF) state. The AF state is stable until the field reverses its direction and increases its value to *H* = –58 Oe, after which $\mathbf{M}_1$ will be aligned with $\mathbf{M}_2$ in the –*y* direction. The forward sweeping process is exactly the opposite process of backward sweeping. When the field strength is reduced to around *H* = –50 Oe, $\mathbf{M}_1$ ($\mathbf{M}_2$) starts to rotate away from the –*y* axis in a clockwise (counterclockwise) manner. After which the second AF state is reached between –34 Oe < *H* < 58 Oe before they saturate again in the +*y* direction. The difference between the first and second AF states is that, in the two states, $\mathbf{M}_1$ and $\mathbf{M}_2$ point in opposite directions. The processes shown in Fig. 7(b) are exactly the same as those of Fig. 7(a) except that the rotation directions of $\mathbf{M}_1$ and $\mathbf{M}_2$ are swopped. Therefore, after the completion of the backward sweeping process, $\varphi_1$ and $\varphi_2$ become 180° and –180°, respectively. Figures 7(a) and (b) are obtained by assuming that the initial sense of rotation for both $\varphi_1$ and $\varphi_2$ is the same for both backward and forward sweeping. However, from energetics point of view, there is no reason that $\varphi_1$ and $\varphi_2$ must follow the same sense of rotation during backward and forward sweeping. Therefore, we still have the other two possible reversal processes. Figure 7(c) shows the case that the backward sweeping follows the same sense of rotation of Fig. 7(b), but forward sweeping is via an opposite sense of rotation, i.e., instead of rotating clockwise (counterclockwise) (Fig. 7(j)), $\varphi_1$ ($\varphi_2$) rotates counterclockwise (clockwise) at the initial stage of magnetization reversal (Fig. 7(k)). Similarly, Fig. 7(d) corresponds to the case that the backward sweeping has the same sense of rotation of Fig. 7(a), but forward sweeping is different (Fig. 7(l)). The same simulations are also conducted for the CoFeB SAF emitters with the parameters: $M_1 = 380$ emu/cm$^3$, $M_2 = 300$ emu/cm$^3$, $K_{u1} = 1 \times 10^4$ erg/cm$^3$, $K_{u2} = 7.75 \times 10^3$ erg/cm$^3$, $d_1 = 4$ nm, $d_2 = 3$ nm, $J_1 = 0.122$ erg/cm$^2$, and $J_2 = 0.055$ erg/cm$^2$. The values $M_1$ and $M_2$ are obtained from the VSM $M$-$H$ curve and $K_{u1}$, $K_{u2}$ value are estimated from the coercivity of respective single layer films. The initial value of $J_1$ is chosen to be comparable with those reported in other works [32-34]. $J_2$ is optimized through simulation. A non-zero value of $J_2$ after fine tuning is used in order to fit the measured VSM results. Figures 7(e)-(h) are the simulated $\varphi_1$ and $\varphi_2$ values as a function of $H$. The results are qualitatively the same as those of FeMnPt SAF except that there is an additional magnetization flipping process shortly after the external field changes directions. This is due to the difference in Zeeman energy which favors the alignment of the FM layer with larger magnetization in the external field direction. Besides this, the large $J_1$ and $J_2$ values also result in a more gradual rotation of $\mathbf{M}_1$ and $\mathbf{M}_2$ compared to the FeMnPt case.

TABLE II. Four different scenarios of rotation directions of $\mathbf{M}_1$ and $\mathbf{M}_2$ subjecting to an external sweeping field (CW: clockwise; CCW: counterclockwise).

|  | Scenario I | Scenario II | Scenario III | Scenario IV |
| --- | --- | --- | --- | --- |
| Backward sweeping | $\mathbf{M}_1$: CW <br> $\mathbf{M}_2$: CCW | $\mathbf{M}_1$: CCW <br> $\mathbf{M}_2$: CW | $\mathbf{M}_1$: CCW <br> $\mathbf{M}_2$: CW | $\mathbf{M}_1$: CW <br> $\mathbf{M}_2$: CCW |
| Forward sweeping | $\mathbf{M}_1$: CW <br> $\mathbf{M}_2$: CCW | $\mathbf{M}_1$: CCW <br> $\mathbf{M}_2$: CW | $\mathbf{M}_1$: CW <br> $\mathbf{M}_2$: CCW | $\mathbf{M}_1$: CCW <br> $\mathbf{M}_2$: CW |

## APPENDIX B: GENERATION OF SUPERDIFFUSIVE BACKFLOW CURRENTS AND SUBSEQUENT TERAHERTZ EMISSION

As we explained in our previous work [19], when a MgO/FM/quartz trilayer (here, FM refers to metallic ferromagnet) is irradiated by an fs laser, the electrons are excited to states above the Fermi level in the FM layer (Fig. 8(a)). Immediately after the excitation, equilibration takes place through two dominant mechanisms, i.e., electron-electron and electron-phonon interactions. Due to the much smaller heat capacity, the electron subsystem quickly reaches a high temperature ($T_e$) Fermi-Dirac distribution within 0.1 ps (Fig. 8(b)), whereas the lattice stays close to the ambient temperature ($T_p$). The electrons subsequently cool and thermalize with its own lattice within a few picoseconds (Fig. 8(c)). In addition to phonons, the magnon also plays a role in the equilibration process of hot electrons in ferromagnets. The magnon temperature ($T_m$) is typically higher than that of phonons before all the three subsystems reach the thermal equilibrium [19].

Before the electron subsystem reaches the equilibrium (i.e., $t < 0.1$ ps), the nonthermal electrons move at a fast speed (~$10^6$ m/s) in a super-diffusive manner (Fig. 8(d)) [19]. For electrons at a distance of at least a mean-free-path ($\lambda_e$) away from the top (e.g., with MgO capping layer) and bottom (e.g., with quartz substrate) interfaces, they will collide with other electrons to reach the thermal equilibrium within the electron subsystem. However, for electrons with a distance shorter than $\lambda_e$ from the interface, they will be reflected back to form a backflow current due to reflection at the FM/dielectric interface. The amplitude of the backflow current depends strongly on the material properties and roughness of the interface. For metallic films deposited on a smooth substrate, typically the bottom interface will be smoother than the top interface. In this case, the backflow currents from the bottom interface ($j_{l1}$) is larger than that from the top interface ($j_{l2}$). Therefore, a net longitudinal current is formed in the FM layer, which is subsequently converted to a transverse transient current via anomalous Hall effect (AHE), thereby generating the THz emission.

The same process is expected to occur in the SAF structure as well. As illustrated in Fig. 8(d), in this case, the backflow current from the quartz substrate and MgO capping layer generates a spin polarized current in $\mathbf{M}_1$ and $\mathbf{M}_2$, respectively. The contribution of backflow current from the Ru/FM interface is presumably small considering that both layers are metals. The net longitudinal current in each layer is thus mainly determined by the FM/dielectric interface, which is subsequently converted to THz emission via the AHE. This is a reasonable assumption because we have previously estimated that the electron spin diffusion length in FeMnPt is around 1.4 nm [19]. In the FeMnPt SAF of present study, the individual FeMnPt layer thickness is 4 nm, which makes the backflow spin current hard to reach and cross the FeMnPt/Ru interface. Unlike the single layer FM case, in which the THz emission due to the backflow currents from the top and bottom interfaces cancel each other partially, here the THz emissions from the two layers simply add up due to the antiparallel alignment of the two FM layers at zero field. This is the reason by the SAF structure generates a much stronger THz emission at the same equivalent thickness. In addition to the backflow current from the FM/dielectric interface and subsequent THz generation via AHE inside the FM layer, one should also consider the contribution from the backflow current at the FM/Ru interface and the role of Ru as spin to charge converter, which are both insignificant as elaborated below.

A small reflection of superdiffusive current at the FM/Ru interfaces also means that the superdiffusive current generated inside each FM layer would enter the Ru directly. But, we have enough reason to believe that the currents from the two FM layers are mostly cancelled out inside the Ru layer due to their opposite directions. Even if they are not completely cancelled out, we argue that the effect should be very small. The spin-orbit interaction in Ru is about 3 times smaller than that in Pt [21], and the same relation also holds for the spin Hall angle, 0.04 for Ru and ~0.11 for Pt [22]. As the amplitude of THz emission from FM/non-magnetic metal (NM) bilayers is found to be proportional to the spin Hall angle of the NM layer [2], the THz emission from Ru, if any, should be about 3 times smaller than that from the FeMnPt layer as the anomalous Hall angle of FeMnPt is comparable to that of Pt (assuming the laser-induced superdiffusive current is the same). Of course, one should be mindful of these values as even the spin Hall angle for Pt, the best studied material, varies in a large range. Put these values aside, experimentally, Zhang *et al.* found that THz emission from Ru-based heterostructure is much smaller than those from the same structure but with Ru being replaced by Pd [12]. The two factors combined, i.e., very small net current and small spin Hall angle, ensure that the contribution from Ru to the THz emission can be safely ignored.

Another factor which may affect the THz emission is the "spilling" over of superdiffusive current across the Ru layer entering the other side of the FM. This is in fact a desirable effect for the FeMnPt SAF. As we reported previously [19], the THz emission in single layer FM emitter originates from the backflow current from the top and bottom interfaces. As the reflected currents are always in opposite directions, the contributions from the two surfaces partially cancel out with each other, and the net emission is the difference between the two contributions. In order to increase the net emission, we had to increase the difference in contributions from the two interfaces. This was the reason why in the previous work we introduced a composition gradient and, in this work, we introduced the SAF structure. The superdiffusive current enters FeMnPt layer at the other side should be smaller than the backflow current from the FM/oxide interface in a single layer structure. Therefore, if we divide the SAF into two halves, the THz emission from each half should be larger than that of the equivalent single layer structure. In fact, this is exactly what we have observed.

**FIGURE CAPTIONS**

FIG. 1. Schematic of the (a) FeMnPt SAF and (b) CoFeB SAF structures used in this study. (c) Schematic of the coordinate systems used for THz detection.

FIG. 2. Simulated longitudinal (along $y$-axis) and transverse (along $x$-axis) components of total (i.e., $\mathbf{M} = \mathbf{M}_1 + \mathbf{M}_2$) and differential (i.e., $\Delta\mathbf{M} = \mathbf{M}_1 - \mathbf{M}_2$) components of the magnetization of a FeMnPt SAF versus a sweeping external field ($H$). (a)-(d), $M_y = M_{y1} + M_{y2}$; (e)-(h), $\Delta M_y = M_{y1} - M_{y2}$; (i)-(l), $-\Delta M_x = M_{x2} - M_{x1}$. The four different cases in each row correspond to the four different combinations of $\mathbf{M}_1$ and $\mathbf{M}_2$ rotation directions when the SAF is subjected to the sweeping field, as detailed in Appendix A. Solid-lines and dashed-lines represent the forward and backward sweeping, respectively. Blue and red arrows indicate $\mathbf{M}_1$ and $\mathbf{M}_2$ directions (right: +$y$ direction; left: –$y$ direction).

FIG. 3. Simulated longitudinal (along $y$-axis) and transverse (along $x$-axis) components of total (i.e., $\mathbf{M} = \mathbf{M}_1 + \mathbf{M}_2$) and differential (i.e., $\Delta\mathbf{M} = \mathbf{M}_1 - \mathbf{M}_2$) components of the magnetization of a CoFeB SAF versus a sweeping external field ($H$). (a)-(d), $M_y = M_{y1} + M_{y2}$; (e)-(h), $-\Delta M_y = M_{y2} - M_{y1}$; (i)-(l), $\Delta M_x = M_{x1} - M_{x2}$. The four different cases in each row correspond to the four different combinations of $\mathbf{M}_1$ and $\mathbf{M}_2$ rotation directions when the SAF is subjected to the sweeping field, as detailed in Appendix A. Solid-lines and dashed-lines represent the forward and backward sweeping, respectively. Blue and red arrows indicate $\mathbf{M}_1$ and $\mathbf{M}_2$ directions (right: +$y$ direction; left: –$y$ direction).

FIG. 4. Magnetic properties of the SAFs. The solid-lines represent forward sweeping and dashed-

lines represent backward sweeping. (a) *M-H* curves of the ($d_1$, $d_2$) FeMnPt SAFs. (b) Temperature dependence of the *M-H* loop of the (4, 4) FeMnPt SAF. (c) *M-H* curves of the ($d_1$, $d_2$) CoFeB SAFs. Inset: *M-H* curves in the field region of –200 Oe to 200 Oe. (d) Temperature dependence of the *M-H* loops of the (4, 3) CoFeB SAF. Inset: *M-H* curves in field region of –100 Oe to 100 Oe.

FIG. 5. Time domain THz emission waveforms from the SAF structures. (a) THz waveforms of the (4, 4) FeMnPt SAF at 200 Oe (dashed-line) and 0 Oe (solid-line), and FeMnPt (5) single layer emitter (dotted-line) at a positive magnetic field. (b) THz waveforms of the (4, 3) CoFeB SAF at 2000 Oe (dashed-line) and 0 Oe (solid-line), and CoFeB(3)/Pt(3) bilayer emitter (dotted-line) at a negative magnetic field. (c) Pump fluence dependence of the THz peak amplitude for the FeMnPt (diamond) and CoFeB (square) SAFs.

FIG. 6. Magnetic field dependence of the THz emission. The solid-lines represent the forward sweeping and dashed-lines represent backward sweeping. (a) $E_x$-*H* curve of the (4, 4) FeMnPt SAF at pumping fluences from 110 to 385 μJ/cm$^2$; (b) $E_y$-*H* curve of the (4, 4) FeMnPt SAF at a pumping fluence of 330 μJ/cm$^2$; (c) $E_x$-*H* curve of the (4, 3) CoFeB SAF structures at fluences from 110 to 550 μJ/cm$^2$; (d) $E_y$-*H* curve of the (4, 3) CoFeB SAF at a pumping fluence of 330 μJ/cm$^2$.

FIG. 7. Simulated $\varphi_1$ and $\varphi_2$ values for (a)-(d) FeMnPt SAF and (e)-(h) CoFeB SAF under a sweeping magnetic field. The four sets of results correspond to the four different combinations of **M**$_1$ and **M**$_2$ rotation directions summarized in Table II. $\varphi_1$ and $\varphi_2$ are positive for counterclockwise rotation and negative for clockwise rotation. The solid-lines represent forward sweeping and dashed-lines represent backward sweeping. (i)-(l) Schematics of the initial magnetization rotation direction during field sweeping for the four different cases (upper half: backward sweeping, lower half: forward

sweeping).

FIG. 8. (a)-(c) Sub-processes at different time scale after ultrafast laser excitation of a metallic layer: (a) Excitation of non-equilibrium electrons. (b) High-temperature Fermi-Dirac distribution of electron subsystem. (c) Nearly thermal equilibrium state among electron, phonon and magnon. (d) Schematic of nonthermal electron reflection at quartz/FM and FM/MgO interfaces, the directions of spin polarity in each FeMnPt layer, the formation of backflow currents, and the charge current directions.

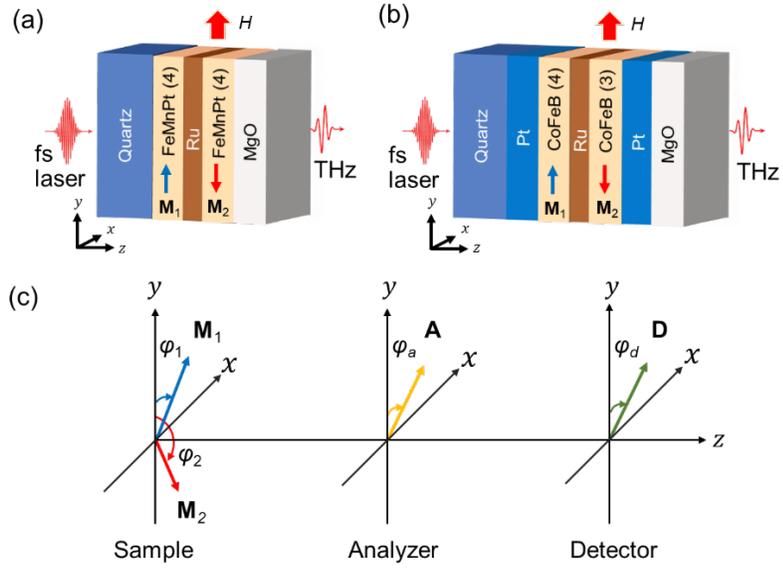

FIG. 1. Schematic of the (a) FeMnPt SAF and (b) CoFeB SAF structures used in this study. (c) Schematic of the coordinate systems used for THz detection.

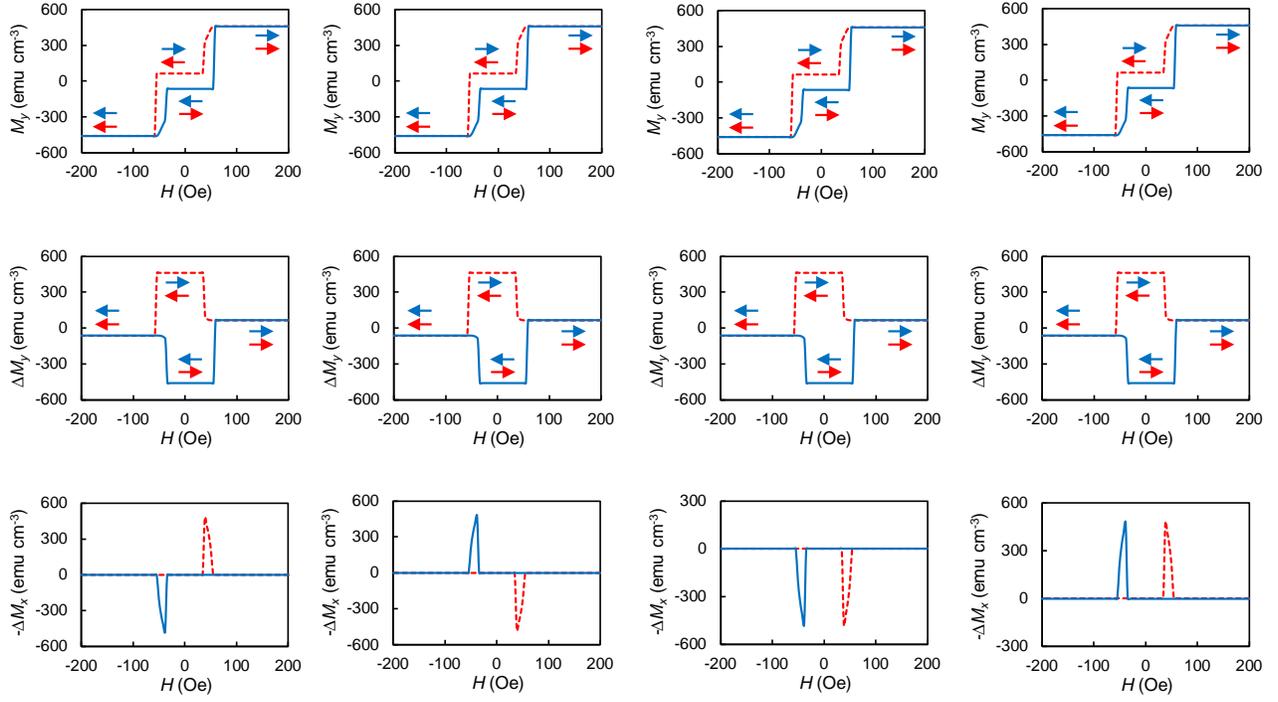

FIG. 2. Simulated longitudinal (along $y$-axis) and transverse (along $x$-axis) components of total (i.e., $\mathbf{M} = \mathbf{M}_1 + \mathbf{M}_2$) and differential (i.e., $\Delta\mathbf{M} = \mathbf{M}_1 - \mathbf{M}_2$) components of the magnetization of a FeMnPt SAF versus a sweeping external field ($H$). (a)-(d), $M_y = M_{y1} + M_{y2}$; (e)-(h), $\Delta M_y = M_{y1} - M_{y2}$; (i)-(l), $-\Delta M_x = M_{x2} - M_{x1}$. The four different cases in each row correspond to the four different combinations of $\mathbf{M}_1$ and $\mathbf{M}_2$ rotation directions when the SAF is subjected to the sweeping field, as detailed in Appendix A. Solid-lines and dashed-lines represent the forward and backward sweeping, respectively. Blue and red arrows indicate $\mathbf{M}_1$ and $\mathbf{M}_2$ directions (right: $+y$ direction; left: $-y$ direction).

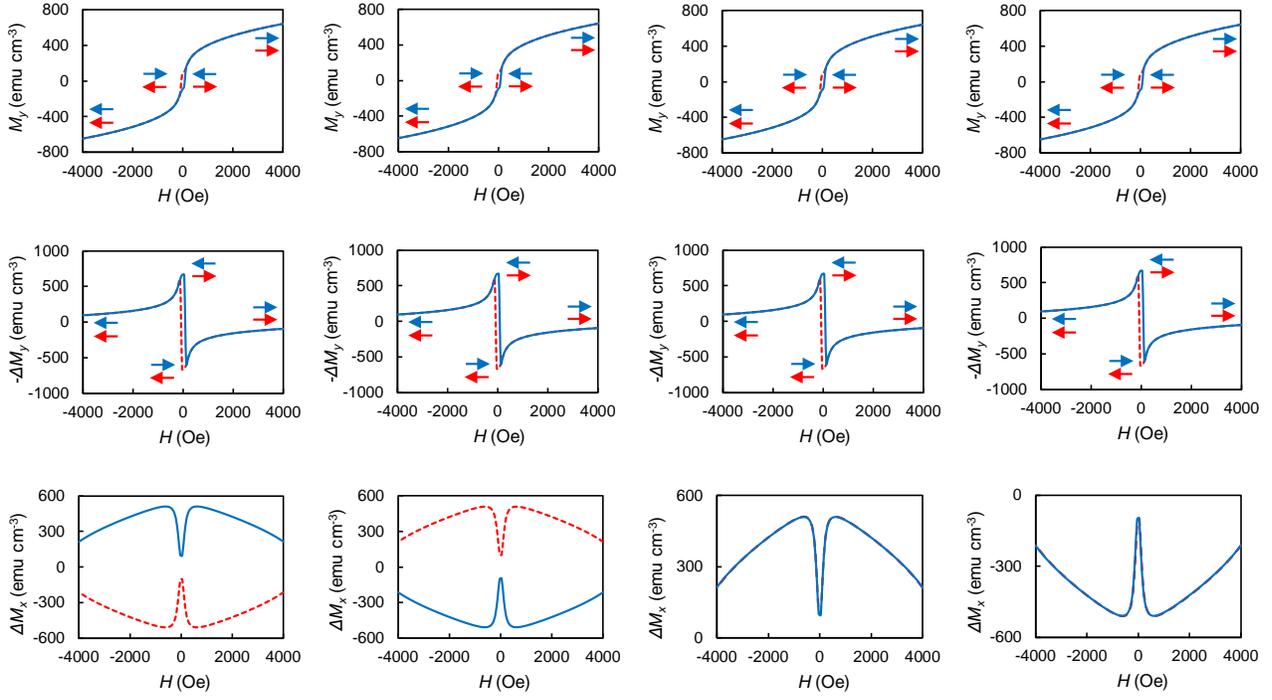

FIG. 3. Simulated longitudinal (along $y$-axis) and transverse (along $x$-axis) components of total (i.e., $\mathbf{M} = \mathbf{M}_1 + \mathbf{M}_2$) and differential (i.e., $\Delta\mathbf{M} = \mathbf{M}_1 - \mathbf{M}_2$) components of the magnetization of a CoFeB SAF versus a sweeping external field ($H$). (a)-(d), $M_y = M_{y1} + M_{y2}$; (e)-(h), $-\Delta M_y = M_{y2} - M_{y1}$; (i)-(l), $\Delta M_x = M_{x1} - M_{x2}$. The four different cases in each row correspond to the four different combinations of $\mathbf{M}_1$ and $\mathbf{M}_2$ rotation directions when the SAF is subjected to the sweeping field, as detailed in Appendix A. Solid-lines and dashed-lines represent the forward and backward sweeping, respectively. Blue and red arrows indicate $\mathbf{M}_1$ and $\mathbf{M}_2$ directions (right: $+y$ direction; left: $-y$ direction).

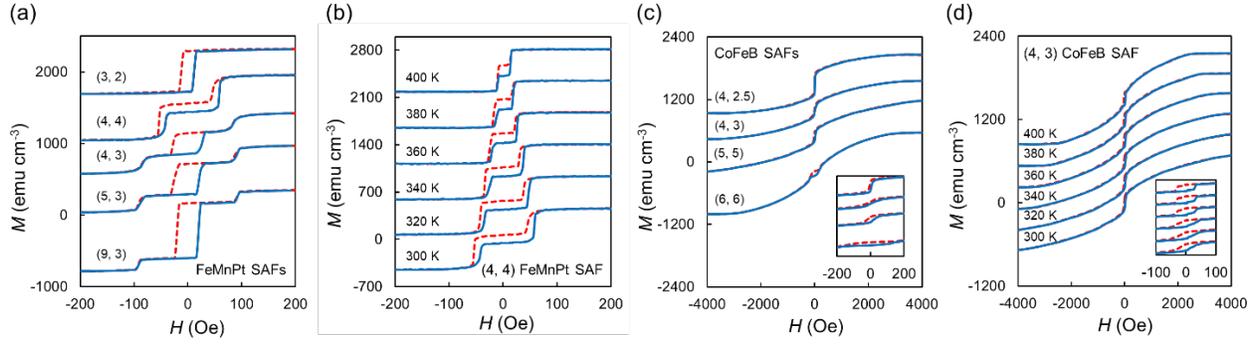

FIG. 4. Magnetic properties of the SAFs. The solid-lines represent forward sweeping and dashed-lines represent backward sweeping. (a) $M$-$H$ curves of the ($d_1$, $d_2$) FeMnPt SAFs. (b) Temperature dependence of the $M$-$H$ loop of the (4, 4) FeMnPt SAF. (c) $M$-$H$ curves of the ($d_1$, $d_2$) CoFeB SAFs. Inset: $M$-$H$ curves in the field region of –200 Oe to 200 Oe. (d) Temperature dependence of the $M$-$H$ loops of the (4, 3) CoFeB SAF. Inset: $M$-$H$ curves in field region of –100 Oe to 100 Oe.

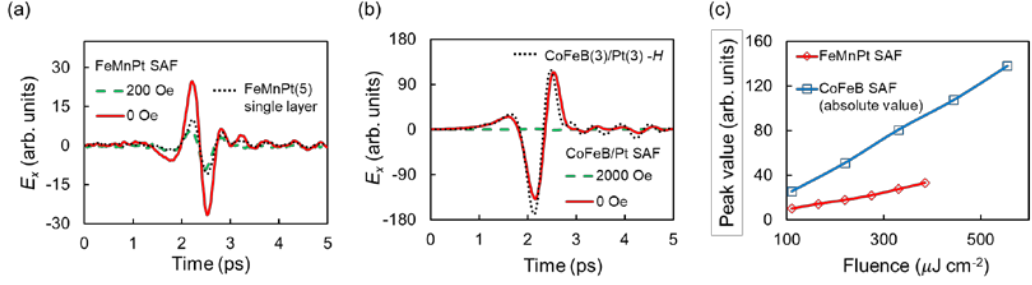

FIG. 5. Time domain THz emission waveforms from the SAF structures. (a) THz waveforms of the (4, 4) FeMnPt SAF at 200 Oe (dashed-line) and 0 Oe (solid-line), and FeMnPt (5) single layer emitter (dotted-line) at a positive magnetic field. (b) THz waveforms of the (4, 3) CoFeB SAF at 2000 Oe (dashed-line) and 0 Oe (solid-line), and CoFeB(3)/Pt(3) bilayer emitter (dotted-line) at a negative magnetic field. (c) Pump fluence dependence of the THz peak amplitude for the FeMnPt (diamond) and CoFeB (square) SAFs.

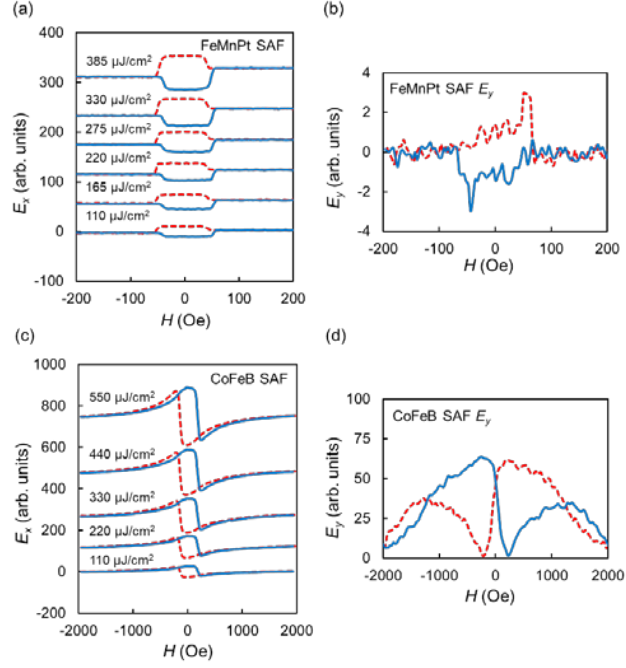

FIG. 6. Magnetic field dependence of the THz emission. The solid-lines represent the forward sweeping and dashed-lines represent backward sweeping. (a) $E_x$-$H$ curve of the (4, 4) FeMnPt SAF at pumping fluences from 110 to 385 μJ/cm$^2$; (b) $E_y$-$H$ curve of the (4, 4) FeMnPt SAF at a pumping fluence of 330 μJ/cm$^2$; (c) $E_x$-$H$ curve of the (4, 3) CoFeB SAF structures at fluences from 110 to 550 μJ/cm$^2$; (d) $E_y$-$H$ curve of the (4, 3) CoFeB SAF at a pumping fluence of 330 μJ/cm$^2$.

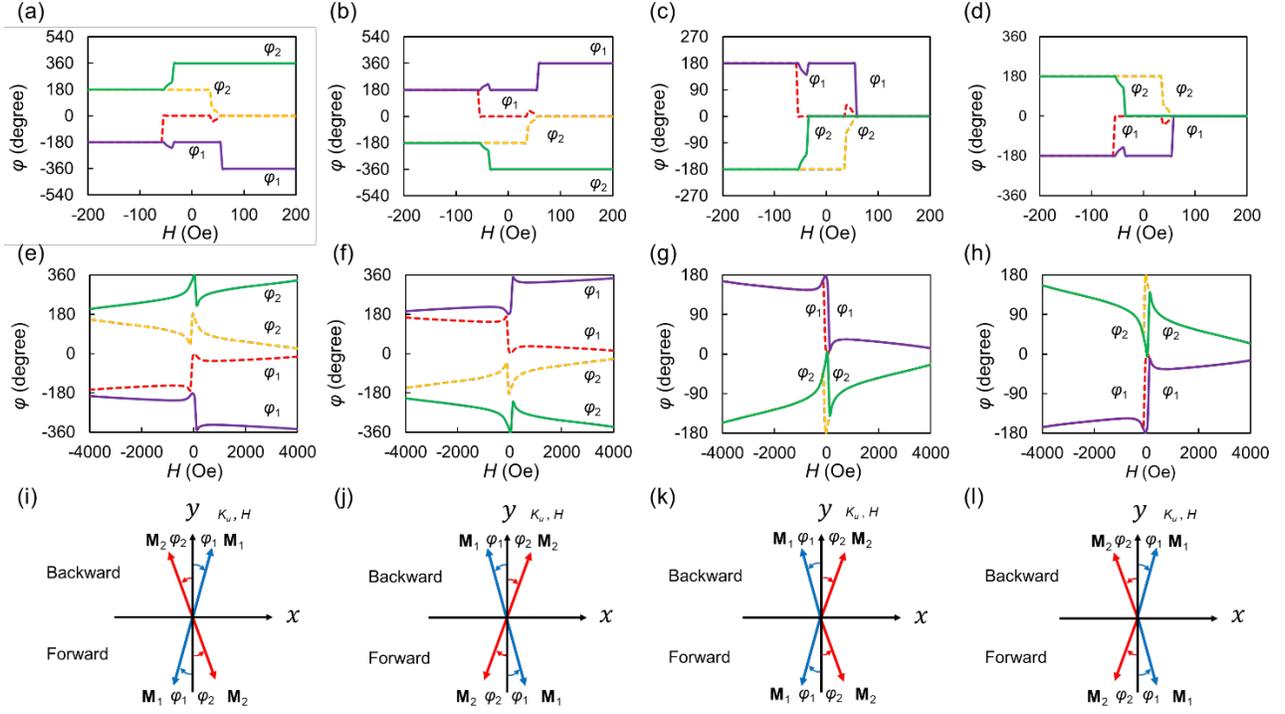

FIG. 7. Simulated $\varphi_1$ and $\varphi_2$ values for (a)-(d) FeMnPt SAF and (e)-(h) CoFeB SAF under a sweeping magnetic field. The four sets of results correspond to the four different combinations of $\mathbf{M}_1$ and $\mathbf{M}_2$ rotation directions summarized in Table II. $\varphi_1$ and $\varphi_2$ are positive for counterclockwise rotation and negative for clockwise rotation. The solid-lines represent forward sweeping and dashed-lines represent backward sweeping. (i)-(l) Schematics of the initial magnetization rotation direction during field sweeping for the four different cases (upper half: backward sweeping, lower half: forward sweeping).

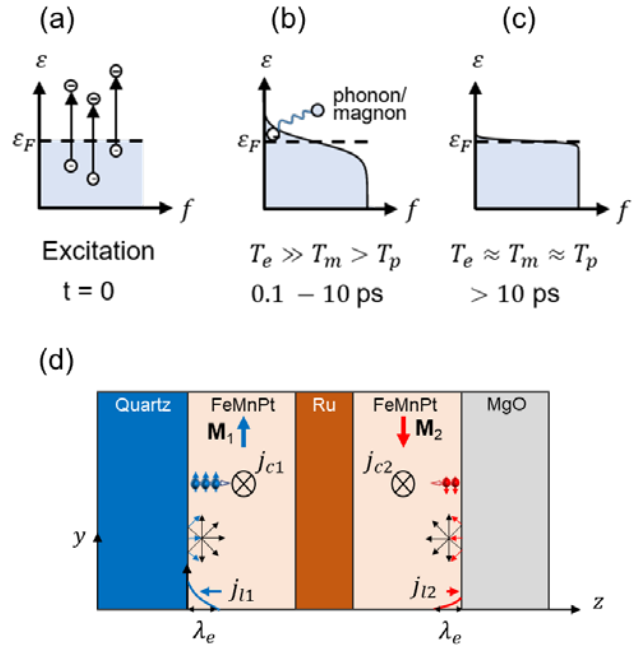

FIG. 8. (a)-(c) Sub-processes at different time scale after ultrafast laser excitation of a metallic layer: (a) Excitation of non-equilibrium electrons. (b) High-temperature Fermi-Dirac distribution of electron subsystem. (c) Nearly thermal equilibrium state among electron, phonon and magnon. (d) Schematic of nonthermal electron reflection at quartz/FM and FM/MgO interfaces, the directions of spin polarity in each FeMnPt layer, the formation of backflow currents, and the charge current directions.